# Simulating the Software Development Lifecycle: The Waterfall Model

Antonios Saravanos [1,*] and Matthew X. Curinga [2]

[1] Division of Applied Undergraduate Studies, New York University, 7 East 12th St, Room 625B, New York, NY 10003, USA
[2] MIXI Institute for STEM and the Imagination, Adelphi University, 179 Livingston St, Brooklyn, NY 11201, USA; mcuringa@adelphi.edu
[*] Correspondence: saravanos@nyu.edu; Tel.: +1-212-992-8725

**Abstract:** This study employs a simulation-based approach, adapting the waterfall model, to provide estimates for software project and individual phase completion times. Additionally, it pinpoints potential efficiency issues stemming from suboptimal resource levels. We implement our software development lifecycle simulation using SimPy, a Python discrete-event simulation framework. Our model is executed within the context of a software house on 100 projects of varying sizes examining two scenarios. The first provides insight based on an initial set of resources, which reveals the presence of resource bottlenecks, particularly a shortage of programmers for the implementation phase. The second scenario uses a level of resources that would achieve zero-wait time, identified using a stepwise algorithm. The findings illustrate the advantage of using simulations as a safe and effective way to experiment and plan for software development projects. Such simulations allow those managing software development projects to make accurate, evidence-based projections as to phase and project completion times as well as explore the interplay with resources.

## 1. Introduction

Despite decades of study and advancements in processes and procedures, software development remains a fraught and challenging process plagued by cost overruns, missed deadlines, abandonment of final products, and outright failure. Research shows a significant probability of failure [1,2]: Bloch et al. [3] share from "a study of 5400 large IT projects exceeding $15 million found that, on average, large IT projects run 45% over budget and 7% behind schedule while delivering 56% less value than initially planned". As Wu et al. [4] point out, information technology (IT) projects are inherently high-risk endeavors—a sentiment echoed since the early 1980s. Charette [5] reiterates this dismal reality when remarking that "few IT projects, in other words, truly succeed", writing that "of the IT projects that are initiated, from 5 to 15 percent will be abandoned before or shortly after delivery". The author characterizes this "as hopelessly inadequate" [5]. Zykov [6] writes, "the product that the developers produce may significantly differ from what the customers expects". Certainly, the reputation of the practice of software development has suffered from its inability to guarantee success. Accordingly, Kellner et al. [7] find that "over the past few decades the software industry has been assailed by numerous accounts of schedule and cost overruns as well as poor product quality delivered by software development organizations in both the commercial and government sectors".

Data on the magnitude of the challenges and failures with respect to software development are traditionally sought from the Standish Group's CHAOS reports. DeFranco and Voas [8] have indicated that the Standish Group has been gathering data about project outcomes (i.e., the successful or failure completion of projects) since 1994 through their CHAOS reports. Morawiec et al. [9] provide specifics from these reports, showing that in 2015, 29% of projects were successful and 19% failed [10]. By 2020, the success rate had increased slightly to 31%, while the failure rate remained at 19% [11]. In other words, almost one in five IT projects will fail. Alves et al. [12] in their review of crisis and risks in engineering project management describe the existence of a software crisis. Middleton and Sutton [13] share that "people who have labored to produce software in chaotic development environments appreciate anything that brings more order and predictability to their work". We find hope in the words of Charette [5], who writes that "the biggest tragedy is that software failure is for the most part predictable and avoidable".

To enhance the chances of success, most firms implement a software development life cycle (SDLC). This can be thought of as a recipe "for designing, building, and maintaining information" [14] systems by prescribing steps "that are essential for developers, such as planning, analysis, design, and implementation" [15]. The innovation emerged in the 1960s after "the failure of several high-profile software projects" [16]. To support the use of a software development lifecycle such as the waterfall model, management can rely on simulations of software process models to support decision making and minimize the risk of failure, as proposed by Kellner et al. [7]. The drive here is that "we use models in an attempt to gain understanding and insights about some aspect of the real world" [17]. Certainly, the use of simulations to better understand a current scenario as well as competing ways one can work with the lifecycle is further supported by Pinho et al. [18], who stress that efficient business management depends in part on the ability to evaluate such competing scenarios. Furthermore, simulation can also be regarded as an effective way of gleaning valuable insights "when the costs, risks or logistics of manipulating the real system of interest are prohibitive" [7]. Vasilecas et al. [19] note that the allocation of resources is a particular area where one can frequently encounter challenges.

*Our Contribution and the Remainder of this Paper*

There have been several attempts to simulate software process models (e.g., [20–23]) which include waterfall (the most cited to our knowledge being Bassil's [21], which we will attempt to align with how we structure our simulation for uniformity). Our contribution is to demonstrate how an open-source programming language such as Python coupled with the SimPy framework [24] can be used to simulate the waterfall model in order to ascertain the optimal number of resources to minimize the possibility of bottlenecks and idle resources, the implications being that it can be more easily used in industry and education to offer valuable insight that could be used by those managing, or learning to manage, software development projects.

The remainder of this paper unfolds as follows: Section 2 delves into the background of the waterfall software development lifecycle and our motivation for its study. In Section 3, we outline the materials and methods used with respect to our simulation. Section 4 presents the results following sample executions of the simulation, demonstrating how it can be used to identify resource levels that would result in zero-wait times. The findings from this exercise are discussed in Section 5. Finally, Section 6 concludes the paper by outlining how this work could be applied as well as the limitations which concurrently offer directions for how this work could be advanced.

**2. Background**

In this section, we provide an overview of the relevant literature introducing the innovation of software development processes and the idea of a lifecycle. We then present the waterfall model, which is perceived as the first software development lifecycle, as well as our motivation for its study.

## 2.1. The Software Development Lifecycle

The SDLC "is also called [the] software development process model" [25], with Sarker et al. [26] equating the two terms. The term "software development life cycle" similarly appears in the literature to be used at times interchangeably with the "systems development life cycle". Ruparelia [27] rationalizes this equation by pointing out that "systems development lifecycle models have drawn heavily on software and so the two terms can be used interchangeably in terms of SDLC". However, in practice, "a software development life cycle" can frequently be "considered as a subset of [the] system development life cycle" [26].

Öztürk [28] reviews several definitions for the term SDLC and distills them to offer his own, which we will use as the basis for this work, specifically, to be "a model that defines the sequence of phases and activities that will/should take place during the software development". The SDLC "covers all stages of software from its inception with requirements definition through to fielding and maintenance". Over time, a number of SDLCs have emerged "which have their own advantages/disadvantages and strengths/weakness" [28]. The inception of the first such model is traced by Dawson and Dawson [29] to the pioneering work of Benington [30]. Benington heralds SDLCs' importance, writing, "since the first software process model was introduced by Benington in 1956, software engineers have continued to recognize that processes are the bedrock on which all software developments succeed".

While SDLCs are not new, they are still evolving along with advances in computing and software engineering. Existing models are being extended to include specific extensions for secure software development [31], green computing [32] that considers the material and environmental impact of software systems, and the impact of machine learning/deep learning both as it provides insight into the development process and presents unique challenges to the SLDC of machine learning systems [33] where requirements are emergent from the analysis of data rather than gathered from stakeholders.

## 2.2. The Waterfall Model

Rastogi [34] presents what is colloquially known as "the waterfall model" as "the classical model of software engineering" as well as "one of the oldest models" [26]. Balaji and Murugaiyan [15] also support this characterization of the waterfall model as "the oldest of these, and the best known". Sommerville [16] describes it as "the initial life-cycle model", "now termed the waterfall model". We should point out that the model is based on and expands on the work of Benington (which was introduced earlier). This waterfall model is classified by Dennis et al. [35] under the category of "structured design", who go on to note that this category replaced "the previous ad hoc and undisciplined approach" that was being used to develop software. Weisert [36] shares that he "can't find a rigorous definition" for the original waterfall model. Also known as the cascade model [27], the waterfall model is described as "a sequential design process, often used in software development processes, in which progress is seen as flowing steadily downwards (like a waterfall) through the phases" [15]. As Sommerville [16] describes it, the "model consists of a set of phases, starting with system specification, with results cascading from one stage to another".

Thus, "the waterfall metaphor was suggested by the inability of water to flow uphill" [36]. The phases run individually one at a time and "all process phases (planning, design, development, testing and deployment) are performed in a sequential series of steps" [22] where the "output of each stage becomes the input for the next" [15]. Cocco et al. [22] elucidate that "each phase starts only when the previous one has ended". There is a bit of debate as to the origins of the label "waterfall" to describe a life cycle approach, with the first use "often attributed either to Barry Boehm or to critics of his COCOMO estimating technique" [36]. However, Petersen et al. [37] point out that, traditionally, it is credited to the work of Winston Royce [38]. Indeed, colloquially the development of what is referred to as the waterfall model is attributed to his article entitled "Managing the Development of Large Software Systems", which introduced the model in Figure 3 of that paper, although the word waterfall does not explicitly appear in his text. The earliest use of "waterfall" to define Royce's [38] model that we could identify was in the 1976 work of Bell and Thayer [39], where they write, "[Royce] introduced the concept of the "waterfall" of development activities".

In Royce's [38] original work, the model is presented as having seven phases: "Systems Requirements", "Software Requirements", "Analysis", "Program Design", "Coding", "Testing", and "Operations". However, many adaptations can be found in the literature (for example, in Petersen et al. [37] and Paul et al. [40]) that modify the exact delineation and description of phases. The waterfall model has received considerable criticism regarding its suitability for development. Weisert [36] points out that "the key attribute of the so-called 'waterfall approach' seems to be extreme inflexibility", explaining that "once you've completed a phase, its results are frozen. You can't go back and revise anything based on changing needs or fresh insights". The author reiterates this point, writing, "until the results of the current phase are complete and approved, you may not start on any work that properly belongs to the next phase or any later phase" [36].

*2.3. Motivation to Study the Waterfall Model*

Despite the growth in popularity of iterative, incremental, and agile software development models, we must recognize the waterfall model's continued relevance and, consequently, its continued need for study. Humphrey and Kellner point out that [41] "outside the research community, much software process thinking is still based on the waterfall framework". This view is echoed by Petersen et al. [37], who write that "waterfall development is still a widely used way of working in software development companies". A study by Andrei et al. [42] in 2019 looked at the usage of agile versus waterfall approaches, finding that 28.1% of software developers had reported that they used waterfall. However, this estimation may be too low—according to the PMI [43] in 2020, 56% of projects used a traditional project management methodology (i.e., waterfall). Fagarasan et al. [44] reinforce the persistence of waterfall, finding that "although the Agile methodology started to become the standard methodology in software projects implementation, several organizations are currently employing Waterfall methodology because it simply works, and it has a proven track record". Further, although waterfall exists in hybrid forms and one of many software development lifecycles, because of its clear stages and easy to grasp model, it remains foundational for computing education as novice software engineers learn more about the software development process.

While having a place in software development, some have questioned whether we have taken the new agile approaches too far [45], implying that there may still be some value in waterfall. This type of thinking has led to an emerging practice of taking the best of waterfall and agile approaches, resulting in hybrid models. Kirpitsas and Pachidis [46] explain the motivation for mixing agile and waterfall: "the rise of hybrid software development methods", which "combined elements from both the waterfall and agile methodologies" do so "to increase efficiency throughout the software development lifecycle". For example, Bhavsar et al. [47] propose a hybrid framework integrating Kanban and Scrum approaches with waterfall. Fagarasan et al. [44] propose another hybrid framework that combines the best from both agile and waterfall. Thus, the study of waterfall is appropriate.

Others (e.g., Sommerville [48], Dennis et al. [49], Dennis et al. [35]) take a more agnostic approach to waterfall, settling on a position where there is no one software development lifecycle that is superior to the others; rather, each has its strengths and weaknesses. So, agile is not superior to waterfall or the reverse; rather, each is designed to address a particular situation. Waterfall appears best suited for projects where requirements need to be well understood upfront and then are not very likely to significantly change over the course of the project [48]. This would make the waterfall approach a sensible option for systems that are complex and need to be reliable [49] (e.g., safety- or security-critical systems [48]).

However, the waterfall model may be best known for its weaknesses. Royce [38] received considerable criticism for the model, even though he understood (perhaps only in part) the weakness of his model. He himself pointed out the problems, writing, "I believe in this concept, but the implementation described above is risky and invites failure" [38]. Bassil [21] emphasizes that the "SDLC of software systems has always encountered problems and limitations that resulted in significant budget overruns, late or suspended deliveries, and dissatisfied clients". He goes on to explain that "the major reason for these deficiencies is that project directors are not wisely assigning the required number of workers and resources on the various activities of the SDLC". As a result, "SDLC phases with insufficient resources may be delayed; while, others with excess resources may be idled, leading to a bottleneck between the arrival and delivery of projects and to a failure in delivering an operational product on time and within budget" [21]. It is these problems that we seek to address using simulation.

## 3. Materials and Methods

Software process simulation and visualization is critical for helping us understand the risks and indicators of failure in software projects. Tracing its roots back to the 1980s, as noted by Zhang et al. [50], the technique has evolved considerably over the years. Ruiz et al. [51] define a simulation model as "a computational model that represents an abstraction or a simplified representation of a complex dynamic system". This definition resonates with Acuna et al. [52], who assert that "a software process model is an abstract representation of the architecture, design or definition of the software process" [52]. For further context, we can look to Kellner et al. [7], who offer a definition for "process", which they describe as "a logical structure of people, technology and practices that are organized into work activities designed to transform information, materials and energy into specified end result(s)" [53]. There are several adaptations of Royce's [38] original waterfall model. For our study, we will simulate the waterfall software development lifecycle culminating on a model inspired by the work of Bassil [21] and Sommerville [48]. An illustration of our adaptation of the waterfall model can be seen in Figure 1. It comprises five phases: requirements analysis and definition (analysis), system and software design (design), implementation and unit testing (implementation), integration and systems testing (testing), and operations and maintenance (maintenance), which are executed in sequence. The steps in the simulation can be seen in Figure 2. Although the titles of the phases are, for the most part, self-explanatory, their respective definitions can be seen in Table 1.

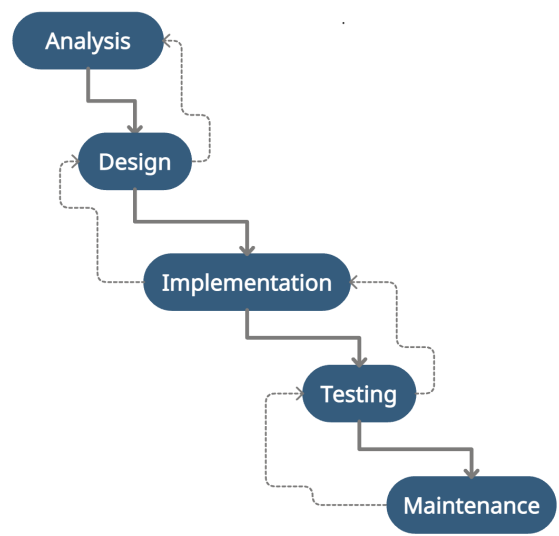

**Figure 1.** Our adaptation of the waterfall model, inspired by Bassil [21] and Sommerville [48].

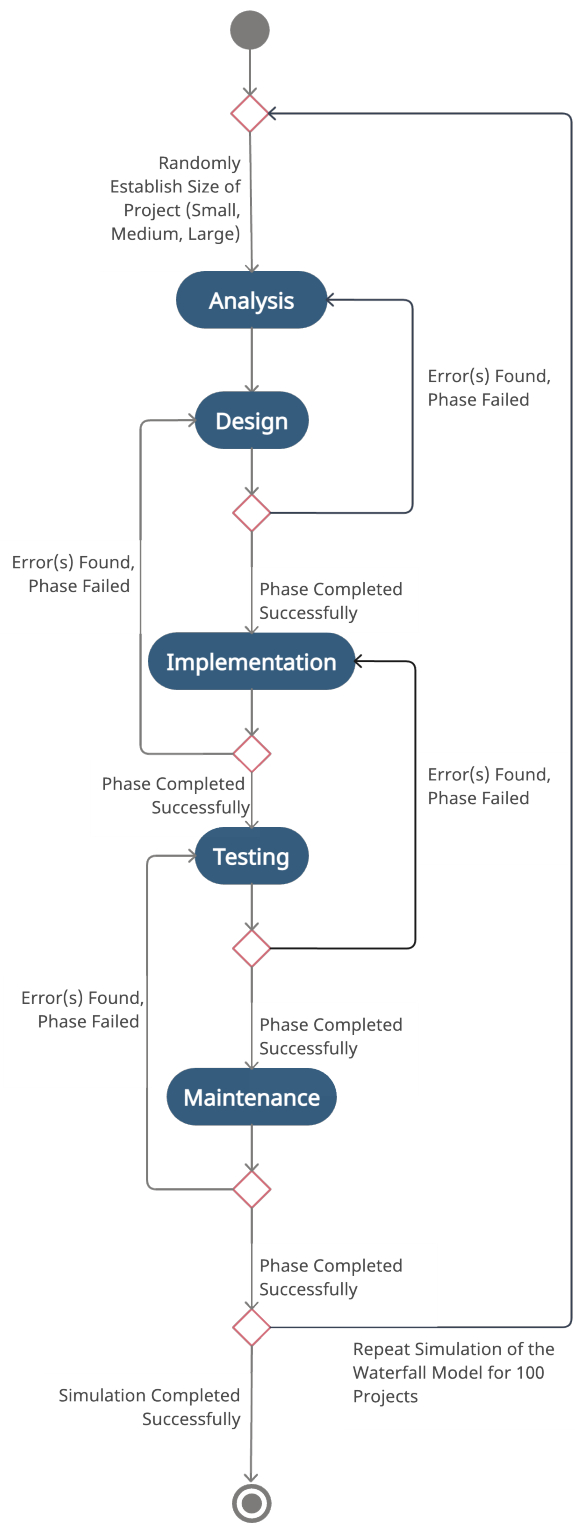

**Figure 2.** Activity diagram illustrating the steps in the simulation.

**Table 1.** List and definition of phases.

| Phase | Definition |
|---|---|
| Requirements **analysis** and definition | "The system's services, constraints, and goals are established by consultation with system users. They are then defined in detail and serve as a system specification" [48]. |
| System and software **design** | "The systems design process allocates the requirements to either hardware or software systems by establishing an overall system architecture. Software design involves identifying and describing the fundamental software system abstractions and their relationships" [48]. |
| **Implementation** and unit testing | "During this stage, the software design is realized as a set of programs or program units. Unit testing involves verifying that each unit meets its specification" [48]. |
| Integration and systems **testing** | "The individual program units or programs are integrated and tested as a complete system to ensure that the software requirements have been met. After testing, the software system is delivered to the customer" [48]. |
| Operation and **maintenance** | "Normally (although not necessarily), this is the longest life cycle phase. The system is installed and put into practical use. Maintenance involves correcting errors which were not discovered in earlier stages of the life cycle, improving the implementation of system units and enhancing the system's services as new requirements are discovered" [48]. |

Note: Phase definitions as stated by Sommerville [48].

*3.1. Assumptions*

Our software simulates a software development house's ability to pursue concurrent projects of small, medium, and large scales, given a fixed number of resources. Given that a simulation is a simplified version of a more complex system [54], we must establish and rely on certain assumptions. To that end, we look to the work of Bassil [21] and his simulation of the waterfall model. Specifically, the available resources for our software house are outlined in Table 2; the required human resources by size of project are presented in Table 3; the duration of phases in days are outlined in Table 4; and the probabilities of an error occurring are presented in Table 5. Note the supposition of clear requirements, which is implicit with the use of the waterfall model.

**Table 2.** Available resources.

| Category | Quantity |
|---|---|
| Analyst(s) | 5 |
| Designer(s) | 5 |
| Programmer(s) | 10 |
| Tester(s) | 20 |
| Maintenance personnel | 5 |

Values are as proposed by Bassil [21].

**Table 3.** Required human resources by scale of project.

| Role | Small | Medium | Large |
|---|---|---|---|
| Analyst(s) | 1 | 2 | 5 |
| Designer(s) | 1 | 2 | 5 |
| Programmer(s) | 2 | 4 | 10 |
| Tester(s) | 2 | 6 | 20 |
| Maintenance personnel | 1 | 2 | 5 |

Values are as proposed by Bassil [21].

**Table 4.** Duration of phases in units of time.

| Phase | Lower | Upper |
|---|---|---|
| Analysis | 3 | 5 |
| Design | 5 | 10 |
| Implementation | 15 | 20 |
| Testing | 5 | 10 |
| Maintenance | 1 | 3 |

Values are as proposed by Bassil [21].

**Table 5.** Possibility of error by scale of project.

| Phase | Small | Medium | Large |
|---|---|---|---|
| Analysis | - | - | - |
| Design | 10% | 20% | 30% |
| Implementation | 10% | 20% | 30% |
| Testing | 10% | 20% | 30% |
| Maintenance | 10% | 20% | 30% |

Values are as proposed by Bassil [21].

We introduce the projects using an exponential statistical function with a mean arrival of 35 units of time (the value of 35 inspired from the work of Bassil [21]) parameterized by a lambda value of 1/35. We also account for the scale of the project, using the work of Jørgensen [55] as guidance. Jørgensen's study of information technology projects reports that 48% of the projects had a budget less than EUR 1 million, 25% between EUR 1 and 10 million, and 27% more than EUR 10 million. Accordingly, we set a 48% chance of a project being a small-scale project, a 25% chance of it being a medium-scale project, and a 27% chance of it being a large-scale project. It should be noted that the software that was developed for this experiment allows for these values to be changed according to the specific needs of the user.

*3.2. Zero-Wait Times*

To identify a resource level which would minimize bottlenecks and result in zero-wait times, we draw upon Riley [56], who provides an overview of existing techniques with respect to discrete-event simulation. The most basic is what is known as "intuitive methods" [56]. Through these, "the user selects input parameters and undertakes an iterative process that involves: (1) varying the parameter levels; (2) completing a statically valid number of simulation replications and runs, and; (3) altering the input parameters and reevaluating the results" [56]. They conclude by noting that "the objective of this method is to find increasingly better solutions" [56]. This is also the approach that we find prescribed in other papers that have attempted to simulate a software development lifecycle (e.g., Bassil [21]). We adopt a similar approach, looking at each resource (i.e., analyst(s), designer(s), programmer(s), tester(s), and maintenance personnel) and associated phase (i.e., requirements analysis, design, implementation, testing, or maintenance) in the waterfall model. Using intuitive methods, we define a minimum and maximum allocation for each resource, as well as an initial step size. An iterative process is then conducted, increasing the corresponding resource until we find the level that results in zero-wait time. This continues until no further improvements can be found based on a provided threshold, where no wait is observed for 3 runs of the simulation.

**4. Results**

In this section, we present the outcomes from simulating our adaptation of the waterfall model. We simulated two scenarios: the first scenario relied on an initial set of resources based on intuition; the second scenario relied on a set of resources identified using a stepwise algorithm to eliminate any bottlenecks caused by a lack of resources. During all executions, we simulated a scenario where 100 projects were started, each of varying size (i.e., small, medium, and large), using the respective probabilities presented in Section 2 and using an exponential statistical function with a mean arrival of 35 units of time. With respect to distribution in Scenario 1, 47% of the projects were small, 31% medium, and 22% large. For Scenario 2, 57% of the projects were small, 20% medium, and 23% large. All initiated projects (i.e., 100) were subsequently completed successfully.

Running the simulation in this first stage provided insight regarding the utilization of resources, specifically allowing us to identify periods where resources were idle as well as periods where there were no available resources, resulting in bottlenecks. Figure 3 illustrates the resource availability and usage from the commencement of the simulation up to the first 400 units of time. We chose to depict only the initial 400 units to maintain clarity in the diagram. Inefficiencies are readily apparent from the data. Specifically, when the line rises above zero, it indicates idle resources. Conversely, a line at zero signifies a resource shortage. The duration for which the line remains at these levels highlights the magnitude of either idleness or shortage. Data from the initial scenario are summarized in Table 6.

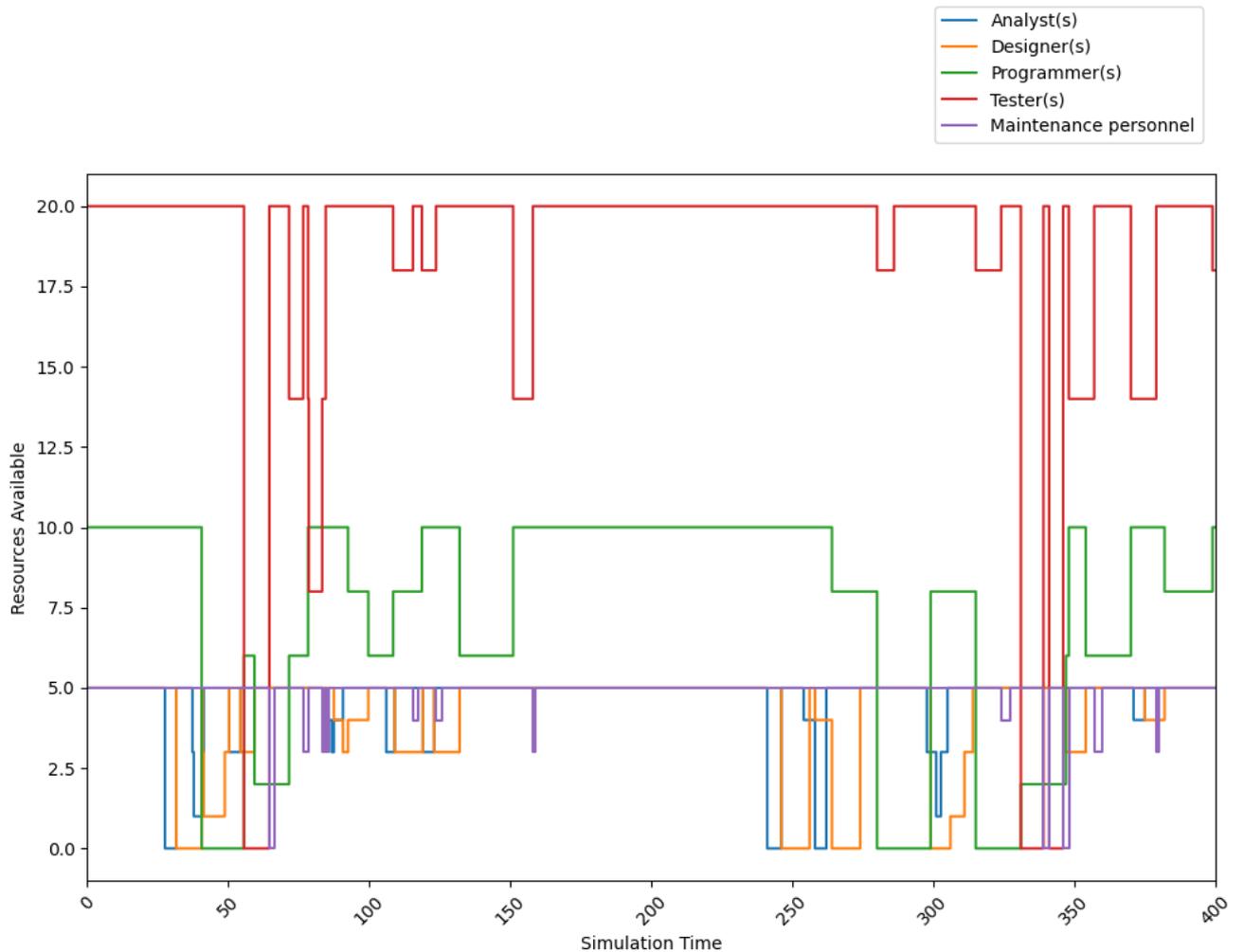

**Figure 3.** Line chart illustrating Scenario 1 utilization of each resource for initial 400 units of simulation time during the first iteration.

Table 6. Count of the number of times that a resource is unavailable for use as well as the mean wait time to acquire that resource in Scenario 1.

| Phase | Resource | Number of Delays | | | Mean Wait Time for Resource | | |
|---|---|---|---|---|---|---|---|
| | | Small | Medium | Large | Small | Medium | Large |
| Analysis | Analyst(s) | 1.000 | 1.000 | 5.000 | 0.010 | 0.069 | 0.388 |
| Design | Designer(s) | 6.000 | 8.000 | 17.000 | 0.525 | 0.380 | 1.030 |
| Implementation | Programmer(s) | 21.000 | 26.000 | 34.000 | 5.344 | 7.422 | 9.707 |
| Testing | Tester(s) | 2.000 | 2.000 | 10.000 | 0.111 | 0.120 | 0.796 |
| Maintenance | Maintenance personnel | 0.000 | 0.000 | 0.000 | 0.000 | 0.000 | 0.000 |
| All phases | All resources | 30.000 | 37.000 | 66.000 | 1.250 | 1.799 | 2.728 |

Note: Units for mean wait time represent simulated units of time.

Under the first scenario, which took 7522.174 units of time to complete, the data revealed that four of the five phases (i.e., analysis, design, implementation, and testing) experienced a wait for their respective resource. The implementation phase, specifically involving programmers, was a pronounced bottleneck. It exhibited the highest mean number of delays across various resource sizes—21.000 for small-sized projects, 26.000 for medium-sized projects, and 34.000 for large projects. Additionally, this phase also displayed the most extended mean wait times, with small projects waiting 5.344 simulated time units for programmers, medium-sized projects 7.422 units of time, and large projects 9.707 units of time. The design phase also faced some bottlenecks, with an overall number of delays of 31.000 and a mean wait time of 0.645 simulated units of time.

On the other hand, the analysis and testing phases demonstrated relatively lower delays and wait times. Small and medium projects each experienced delays in acquiring one or more testers in 2.000 instances, respectively, and large projects experienced delays in 10.000 instances and had to wait on average 0.333 units of time for a tester. The analysis phase had a low overall number of delays of 7.000 and the lowest mean wait time of 0.143 units of time.

Interestingly, the maintenance phase had zero delays and wait times across all resource sizes. Large-sized projects, which required a greater number of the respective resources to initiate their phases, experienced the greatest number of delays as well as the greatest wait times for resources. These findings would strongly advocate for the development of targeted strategies, with an emphasis on the implementation and design phases, and their respective resources, to substantially mitigate delays and decrease wait times across the board.

We ran a second simulation using a stepwise algorithm to identify and set resources to a level that resulted in zero-wait times. Simulation cycles ran recursively until there were no delays for any resources for a certain number of iterations (i.e., three). When a delay occurred, the total number of the resource which was lacking was incremented by one. In total, 40 runs of the simulation were required. For the analysis phase, it was identified that 15 analysts would be needed, in lieu of the initially speculated 5, to address the earlier identified delays.

With respect to the design phase, the designer count was raised from 5 to 18. Successively, we examined the implementation phase; the number of programmers was substantially increased from 10 to 38. The next phase to be examined was testing. We ascertained that the necessary number of testers was 49, rather than the original number of 20 testers. Lastly, the originally proposed level of 5 maintenance personnel for the maintenance phase was increased to 10. These values are summarized in Table 7.

Table 7. Number of original and optimized resources as well as their corresponding phases.

| Phase | Resource | Scenario 1 (Intuition) | Scenario 2 (Zero-Wait) |
|---|---|---|---|
| Analysis | Analyst(s) | 5 | 15 |
| Design | Designer(s) | 5 | 18 |
| Implementation | Programmer(s) | 10 | 38 |
| Testing | Tester(s) | 20 | 49 |
| Maintenance | Maintenance personnel | 5 | 10 |

The second scenario, after adjusting the number of resources in each phase to achieve zero-wait, took 5754.000 units of time to complete. Indeed, we observe that there are no longer any delays to procure respective resources, as can be seen in Table 8, which lists the number of times a delay occurred, as well as the maximum, mean, and standard deviation of wait times for resources across iterations by size of project.

Table 8. The number of times a delay occurred (Count), as well as the maximum (Max Wait), mean (Mean Wait), and standard deviation (Std. Dev.) of wait times, for resources by size of project.

| Phase | | Small | | Medium | | Large | | All Sizes | |
|---|---|---|---|---|---|---|---|---|---|
| | | Scenario 1 | Scenario 2 | Scenario 1 | Scenario 2 | Scenario 1 | Scenario 2 | Scenario 1 | Scenario 2 |
| Analysis | Count | 1.000 | 0.000 | 1.000 | 0.000 | 5.000 | 0.000 | 7.000 | 0.000 |
| | Max Wait | 0.509 | 0.000 | 2.979 | 0.000 | 4.707 | 0.000 | 4.707 | 0.000 |
| | Mean Wait | 0.010 | 0.000 | 0.069 | 0.000 | 0.388 | 0.000 | 0.143 | 0.000 |
| | Std. Dev. | 0.072 | 0.000 | 0.454 | 0.000 | 1.091 | 0.000 | 0.668 | 0.000 |
| Design | Count | 6.000 | 0.000 | 8.000 | 0.000 | 17.000 | 0.000 | 31.000 | 0.000 |
| | Max Wait | 8.977 | 0.000 | 6.000 | 0.000 | 8.059 | 0.000 | 8.977 | 0.000 |
| | Mean Wait | 0.525 | 0.000 | 0.380 | 0.000 | 1.030 | 0.000 | 0.645 | 0.000 |
| | Std. Dev. | 1.754 | 0.000 | 1.124 | 0.000 | 1.989 | 0.000 | 1.669 | 0.000 |
| Implementation | Count | 21.000 | 0.000 | 26.000 | 0.000 | 34.000 | 0.000 | 81.000 | 0.000 |
| | Max Wait | 42.000 | 0.000 | 29.000 | 0.000 | 46.396 | 0.000 | 46.396 | 0.000 |
| | Mean Wait | 5.344 | 0.000 | 7.422 | 0.000 | 9.707 | 0.000 | 7.490 | 0.000 |
| | Std. Dev. | 9.482 | 0.000 | 9.897 | 0.000 | 11.478 | 0.000 | 10.407 | 0.000 |
| Testing | Count | 2.000 | 0.000 | 2.000 | 0.000 | 10.000 | 0.000 | 14.000 | 0.000 |
| | Max Wait | 4.000 | 0.000 | 3.000 | 0.000 | 8.000 | 0.000 | 8.000 | 0.000 |
| | Mean Wait | 0.111 | 0.000 | 0.120 | 0.000 | 0.796 | 0.000 | 0.333 | 0.000 |
| | Std. Dev. | 0.604 | 0.000 | 0.594 | 0.000 | 1.989 | 0.000 | 1.262 | 0.000 |
| Maintenance | Count | 0.000 | 0.000 | 0.000 | 0.000 | 0.000 | 0.000 | 0.000 | 0.000 |
| | Max Wait | 0.000 | 0.000 | 0.000 | 0.000 | 0.000 | 0.000 | 0.000 | 0.000 |
| | Mean Wait | 0.000 | 0.000 | 0.000 | 0.000 | 0.000 | 0.000 | 0.000 | 0.000 |
| | Std. Dev. | 0.000 | 0.000 | 0.000 | 0.000 | 0.000 | 0.000 | 0.000 | 0.000 |
| Across all phases | Count | 30.000 | 0.000 | 37.000 | 0.000 | 66.000 | 0.000 | 133.000 | 0.000 |
| | Max Wait | 42.000 | 0.000 | 29.000 | 0.000 | 46.396 | 0.000 | 46.396 | 0.000 |
| | Mean Wait | 1.250 | 0.000 | 1.799 | 0.000 | 2.728 | 0.000 | 1.901 | 0.000 |
| | Std. Dev. | 4.870 | 0.000 | 5.605 | 0.000 | 6.841 | 0.000 | 5.819 | 0.000 |

Note: The minimum wait times for all resources were 0.000. The units for maximum and mean wait times represent simulated units of time.

The simulation also provided insight regarding project durations; the findings are presented in Table 9 which follows. Overall, we see the fastest project in the second scenario (zero-wait) being completed in 33.000 units of time, this being a large-sized project. The longest project took 168.000 units of time, once again, a large-sized project. The mean completion time for projects for Scenario 2 irrespective of size was 57.540 units of time. Considering project size, for small projects, we see the fastest project completing in 34.000 units of time, the longest project requiring 106.000 units of time, and the mean project completion standing at 47.298 units of time. For medium-sized projects, the quickest was completed in 36.000 units of time and the longest in 112.000 units of time, with the mean at 59.950 units of time. Looking at large-sized projects, we see the fastest project completed in 33.000 units of time, the lengthiest project taking 168.000 units of time, and on average, a project required 80.827 units of time to complete. Therefore, the post-optimization mean time to complete a project increased with the size of the project as it ranged from 47.298 units of time for a small project to 59.950 units of time for a medium project to 80.827 units of time for a large project. We also see the zero-wait algorithm having a mostly positive effect on completion time, as minimum times went from 32.000 to 34.000, from 41.000 to 36.000, from 39.000 to 33.000, and from 32.000 to 33.000 for small, medium, large, and across all sizes, respectively. Similarly, maximum completion times went from 107.141 to 106.000, from 170.142 to 112.000, from 303.000 to 168.000, and from 303.000 to 168.000 for small, medium, large, and across all sizes, respectively. The effects of the optimization process were similar when looking at mean completion times. The values went from 49.995 to 47.298, from 81.944 to 59.950, from 119.643 to 80.827, and from 75.222 to 57.540 for small, medium, large, and across all sizes, respectively. The data clearly indicate that the implementation of a zero-wait strategy can significantly improve project completion times.

**Table 9.** Summary of minimum (Min), maximum (Max), mean, and standard deviation (Std. Dev.) of completion times for projects by size of project for both Scenarios 1 and 2.

|  | Small | | Medium | | Large | | All Sizes | |
|---|---|---|---|---|---|---|---|---|
|  | Scenario | | Scenario | | Scenario | | Scenario | |
|  | 1 | 2 | 1 | 2 | 1 | 2 | 1 | 2 |
| Min | 32.000 | 34.000 | 41.000 | 36.000 | 39.000 | 33.000 | 32.000 | 33.000 |
| Max | 107.141 | 106.000 | 170.142 | 112.000 | 303.000 | 168.000 | 303.000 | 168.000 |
| Mean | 49.995 | 47.298 | 81.944 | 59.950 | 119.643 | 80.827 | 75.222 | 57.540 |
| Std. Dev. | 17.964 | 15.729 | 38.671 | 25.372 | 65.704 | 40.559 | 47.682 | 28.576 |

Note: Units for completion times represent simulated units of time.

We also looked at the duration of the individual phases for both Scenario 1 and Scenario 2. In the analysis phase, we see that the minimum completion time remained a constant 3.000 units across all project sizes, irrespective of scenario. Conversely, the maximum time required for completion in Scenario 2 saw a significant reduction, dropping to 5.000 units from as high as 8.189 units of time for large projects. The mean values showed only minor fluctuations across different project sizes, going from 4.083 units of time for Scenario 1 to 4.058 units of time for Scenario 2, thereby revealing that the size of the project does not significantly impact the average duration in the analysis phase. Moving on to the design phase, the data reveal that while the minimum time remained constant at 5.000 units across all sizes and scenarios, the maximum time required saw a substantial reduction post-optimization. This was most pronounced for large projects, where the maximum time was reduced from 18.059 to 10.000 units of time. Similarly, small projects dropped from 15.541 to 10.000 units of time and medium-sized projects from 15.000 to 10.000 units of time. The average durations showed minimal variation regardless of project size, going from 8.255 to 7.558 units of time. During the implementation phase, the minimum time remained consistent at 15.000 units across all project sizes. However, the maximum time saw a drastic reduction for Scenario 2, dropping to 20.000 units from as high as 62.396 units for large projects. Comparably, for small-sized projects the drop was from 61.000 to 20.000 units of time and for medium-sized projects from 48.000 to 19.000 units of time. The mean times dropped from 24.730 to 17.510 units of time. The testing and maintenance phases showed stable minimum and maximum times for the most part. In the case of the testing phase, a drop was seen with respect to large projects and maximum completion times, which dropped from 16.000 units of time for Scenario 1 to 10.000 units of time for Scenario 2.

In summary, there was a significant reduction in maximum completion times across all phases for Scenario 2. This supports the conclusion that the zero-wait approach was effective in eliminating bottlenecks. Additionally, while large projects generally take more time for Scenario 1, the zero-wait approach effectively narrows this gap, indicating that such a process can improve worst-case scenarios. Table 10 provides greater detail on the duration of phases by size of project across iterations in both scenarios.

**Table 10.** Summary of minimum (Min), maximum (Max), mean, and standard deviation (Std. Dev.) of completion times for phases by size of project for both Scenario 1 and Scenario 2.

| Phase | | Small | | Medium | | Large | | All Sizes | |
|---|---|---|---|---|---|---|---|---|---|
| | | Scenario | | Scenario | | Scenario | | Scenario | |
| | | 1 | 2 | 1 | 2 | 1 | 2 | 1 | 2 |
| Analysis | Min | 3.000 | 3.000 | 3.000 | 3.000 | 3.000 | 3.000 | 3.000 | 3.000 |
| | Max | 5.509 | 5.000 | 6.979 | 5.000 | 8.189 | 5.000 | 8.189 | 5.000 |
| | Mean | 3.890 | 4.095 | 4.093 | 4.083 | 4.313 | 3.975 | 4.083 | 4.058 |
| | Std. Dev. | 0.710 | 0.640 | 0.866 | 0.732 | 1.376 | 0.620 | 1.008 | 0.657 |
| Design | Min | 5.000 | 5.000 | 5.000 | 5.000 | 5.000 | 5.000 | 5.000 | 5.000 |
| | Max | 15.541 | 10.000 | 15.000 | 10.000 | 18.059 | 10.000 | 18.059 | 10.000 |
| | Mean | 7.858 | 7.671 | 8.200 | 7.357 | 8.674 | 7.560 | 8.255 | 7.558 |
| | Std. Dev. | 2.328 | 1.501 | 1.936 | 1.620 | 2.727 | 1.527 | 2.359 | 1.536 |
| Implementation | Min | 15.000 | 15.000 | 15.000 | 15.000 | 15.000 | 15.000 | 15.000 | 15.000 |
| | Max | 61.000 | 20.000 | 48.000 | 19.000 | 62.396 | 20.000 | 62.396 | 20.000 |
| | Mean | 22.398 | 17.458 | 24.668 | 17.207 | 27.125 | 17.760 | 24.730 | 17.510 |
| | Std. Dev. | 9.519 | 1.383 | 9.873 | 1.373 | 11.429 | 1.546 | 10.419 | 1.442 |
| Testing | Min | 5.000 | 5.000 | 5.000 | 5.000 | 5.000 | 5.000 | 5.000 | 5.000 |
| | Max | 10.000 | 10.000 | 10.000 | 10.000 | 16.000 | 10.000 | 16.000 | 10.000 |
| | Mean | 7.278 | 7.433 | 7.540 | 7.500 | 7.980 | 7.320 | 7.588 | 7.406 |
| | Std. Dev. | 1.406 | 1.373 | 1.541 | 1.703 | 2.537 | 1.596 | 1.890 | 1.507 |
| Maintenance | Min | 1.000 | 1.000 | 1.000 | 1.000 | 1.000 | 1.000 | 1.000 | 1.000 |
| | Max | 3.000 | 3.000 | 3.000 | 3.000 | 3.000 | 3.000 | 3.000 | 3.000 |
| | Mean | 2.120 | 1.984 | 1.929 | 2.130 | 1.857 | 2.061 | 1.984 | 2.034 |
| | Std. Dev. | 0.594 | 0.713 | 0.745 | 0.626 | 0.648 | 0.659 | 0.667 | 0.679 |
| Across all phases | Min | 1.000 | 1.000 | 1.000 | 1.000 | 1.000 | 1.000 | 1.000 | 1.000 |
| | Max | 61.000 | 20.000 | 3.000 | 19.000 | 62.396 | 20.000 | 62.396 | 20.000 |
| | Mean | 8.934 | 8.000 | 10.041 | 7.686 | 11.059 | 8.336 | 9.976 | 8.036 |
| | Std. Dev. | 8.539 | 5.497 | 9.523 | 5.125 | 10.804 | 5.602 | 9.654 | 5.449 |

Note: Units for completion times represent simulated units of time.

The simulations also provided insight into the likelihood of a phase failing within the software development lifecycle. It should be noted that, based on our simulation of the waterfall model, errors can first become possible during the design phase. This is in line with the approach proposed by Bassil [21]. Furthermore, we should highlight that the probabilities of an error occurring did not vary between the scenarios, thus having no effect on the possibility of failure. The likelihood of failure was based exclusively on project size. We should also share that our software does allow differentiation by phase should that be desired, namely analysis, design, implementation, testing, and maintenance. The failure rate for small projects was 5.323 percent for Scenario 1 and 7.715 percent for Scenario 2; for medium-sized projects, it was 19.368 percent for Scenario 1 and 17.949 percent for Scenario 2; and lastly, for large projects, the rates were 26.891 percent for Scenario 1 and 24.215 percent for Scenario 2. Overall, the failure rate for Scenario 1 was 16.844 percent, and for Scenario 2, it was 15.084 percent. These data, which detail the number of times a phase failed by project size, are presented in greater detail in Table 11.

**Table 11.** Reports on the mean number of times a phase failed and was not completed successfully by size of project and across iterations.

| Phase | | Small | | Medium | | Large | | All Sizes | |
|---|---|---|---|---|---|---|---|---|---|
| | | Scenario | | Scenario | | Scenario | | Scenario | |
| | | 1 | 2 | 1 | 2 | 1 | 2 | 1 | 2 |
| Design | # of failed phases | 3.000 | 6.000 | 12.000 | 16.000 | 18.000 | 17.000 | 33.000 | 39.000 |
| | # of phases | 54.000 | 73.000 | 61.000 | 42.000 | 59.000 | 50.000 | 174.000 | 165.000 |
| | Percentage | 5.556 | 8.219 | 19.672 | 38.095 | 30.508 | 34.000 | 18.966 | 23.636 |
| Implementation | # of failed phases | 4.000 | 10.000 | 18.000 | 6.000 | 19.000 | 10.000 | 41.000 | 26.000 |
| | # of phases | 55.000 | 72.000 | 57.000 | 29.000 | 55.000 | 50.000 | 167.000 | 151.000 |
| | Percentage | 7.273 | 13.889 | 31.579 | 20.690 | 34.545 | 20.000 | 24.551 | 17.219 |
| Testing | # of failed phases | 4.000 | 5.000 | 8.000 | 3.000 | 14.000 | 17.000 | 26.000 | 25.000 |
| | # of phases | 54.000 | 67.000 | 50.000 | 26.000 | 49.000 | 50.000 | 153.000 | 143.000 |
| | Percentage | 7.407 | 7.463 | 16.000 | 11.538 | 28.571 | 34.000 | 16.993 | 17.483 |
| Maintenance | # of failed phases | 3.000 | 5.000 | 11.000 | 3.000 | 13.000 | 10.000 | 27.000 | 18.000 |
| | # of phases | 50.000 | 62.000 | 42.000 | 23.000 | 35.000 | 33.000 | 127.000 | 118.000 |
| | Percentage (%) | 6.000 | 8.065 | 26.190 | 13.043 | 37.143 | 30.303 | 21.260 | 15.254 |
| All phases | # of failed phases | 14.000 | 26.000 | 49.000 | 28.000 | 64.000 | 54.000 | 127.000 | 108.000 |
| | # of phases | 263.000 | 337.000 | 253.000 | 156.000 | 238.000 | 223.000 | 754.000 | 716.000 |
| | Percentage (%) | 5.323 | 7.715 | 19.368 | 17.949 | 26.891 | 24.215 | 16.844 | 15.084 |

## 5. Discussion

We began this paper inspired by the words of Charette [5], who argued that we could predict and avoid many software failures. Our reliance on software systems continues significant challenges related to exceeding schedules and accordingly costs, as well as the customer's perception of subpar product quality. We feel compelled to develop more robust software development methodologies [51]. Simulation, in particular, has been proposed as a viable solution to challenges encountered in software engineering [57]. In this spirit, we developed and executed an event-driven simulation to augment the software development process. The utilization of software to support the modeling of software projects is a popular exercise, with significant sums of money being spent each year [58]. This simulation, coded in Python using the SimPy framework, is designed to facilitate the generation of accurate estimates of project completion times and identify an optimal allocation of resources, thereby enabling more effective project management. Moreover, it empowers one to understand the interplay and associated tradeoffs that exist between longer projects and procuring more resources. Furthermore, having an accurate forecast of resource requirements can significantly improve financial planning for projects. For instance, exceeding anticipated project timelines can increase project costs and lead to budget overruns. Previous research emphasizes the criticality of delivering high-quality software within resource and time constraints for the software industry [59]. Consequently, it is vital to ensure that project completion is not only within an allocated timeframe but also within a projected budget. Through the use of such simulation, one is empowered to promptly assess alternate scenarios when shifts occur in a planned task timeline in order to realize effective business management [18].

In this work, we built on previous efforts to simulate the waterfall model (e.g., [21,23,60]) by offering an open-source solution utilizing Python and SimPy exploring two scenarios. The first instance simulates the waterfall model with an intuitive set of proposed resources. This provided initial insights into our current resource utilization and identified bottlenecks within the software development life cycle. This suggested a need for increasing resource allocation to ensure sufficient availability of programmers during this phase. We then identify a level of resources that would result in zero-wait. Subsequently, we ran the simulation one final time to ascertain the effect that the optimization process had on the software development life cycle. The results revealed that the adjusted resource levels would achieve no delays. It is important to qualify that this is within a certain margin—running the simulation enough times would theoretically produce at times outliers reflecting resource shortages and phase delays.

The data also offered insight on the errors that could potentially emerge within the software development life cycle. This can offer those managing information systems projects insight as to what errors could appear so that they would not be caught off guard and could adequately address such errors. We would recognize that the presence of errors in the practice of software development is widely recognized. If we were to look at the work of Brooks [61], we would find that approximately half of development time is spent on testing to identify and correct errors. Our analysis revealed that large projects consistently had the highest failure rates, both in Simulation 1 with intuitive resources levels and in subsequent simulations where levels were adjusted to achieve zero-wait.

Kellner et al. [7] write that "a model is valuable to the extent that it provides useful insights, predictions and answers to the questions it is used to address". To that end, this simulator has satisfied this requirement. Certainly, the insight provided would enable anyone managing such an information systems project valuable information to make an informed decision. On the one hand, they would understand and communicate the anticipated duration and predicted completion times for the project and its phases (including any delays caused by a shortage of resources) to the stakeholders. On the other hand, it also offers the potential to explore how changes in resources would influence completion times but also the ability to identify an optimal set of resources that would minimize delays.

## 6. Conclusions

As a practice, the development of software has been thoroughly studied and refined over the years. Despite these efforts, software projects continue to run into trouble or fail entirely. The waterfall model persists as a staple of software development. While there is a belief that waterfall has been discarded in favor of agile approaches, an announcement of its death is premature. A small but significant number of projects still use waterfall. Moreover, there is a growing trend towards hybrid methodologies which combine the more effective aspects of waterfall with the more effective aspects of different flavors of agile.

In our work, we proposed and executed a discrete-event simulation of the waterfall software development life cycle. By proactively identifying and managing potential delays, teams can take preventive measures, better allocate resources, and maintain more predictable project timelines, leading to successful project deliveries and higher customer satisfaction. To implement the simulation, we utilized the Python programming language coupled with the SimPy framework. We executed the simulation based on a set of predefined assumptions which allowed us to estimate the time that would be required to complete various hypothetical projects of different sizes.

The simulation offered insight as to project completion times, resource usage, and any connected impacts on project completion. Furthermore, the simulation assists project managers in determining the resources required to complete a project with minimal bottlenecks and delays, thus providing valuable insights for project managers, software development teams, and organizations seeking to refine and improve their software development processes.

*6.1. Implications*

There are several implications that follow from our efforts, and we would contend that they fall into two broad sets. The first of these is a response to McHaney et al.'s [58] cautioning that "a discrete event computer simulation project can be a complex and difficult undertaking". Our work offers an open-source solution to simulate a software development life cycle using a modern, widely used programming language (i.e., Python). This addresses, in part, the issues of access, cost, and complexity, given that Python, SimPy, and our software are freely available through the internet. Furthermore, we would highlight that the Python language is frequently taught in introductory programming courses. Consequently, many programmers can use our solution without the need for significant retraining. There are few remaining barriers with respect to access, cost, or complexity. Moreover, with respect to complexity, our simulation by design is easily customizable and adaptable. In this way, even those with a rudimentary understanding of Python should be able to use it for their needs. Accordingly, a new population (whether that be industry or academia) should now have access to a solution to simulate the waterfall software development lifecycle.

According to Kellner et al. [7], "we have clustered the many reasons for using simulations of software processes into six categories of purpose: strategic management; planning; control and operational management; process improvement and technology adoption; understanding; and training and learning". Indeed, our solution at the most basic level provides an estimate of a project's duration based on the given set of phases (i.e., analysis, design, implementation, testing, and maintenance) and resources (i.e., in our implementation, business analysts, designers, programmers, testers, and maintenance people). This system allows for better projections for project completion, given an existing collection of resources. It also identifies bottlenecks and provides insight as to the extent of delays and the resources necessary to alleviate or eliminate them. Furthermore, the simulation offers the level of resources necessary to eliminate bottlenecks, thus empowering those managing information systems projects with the ability to strategically plan the lifecycle, understand the interplay of multiple projects, and make any necessary decisions. One way the resource issue could be addressed is by identifying an optimal part-time full-time employment composition to best address the imbalances between idle resources and bottlenecks caused by a lack of resources, always keeping in mind the words of Brooks [61] that adding resources takes time.

Not only is such a simulation valuable for the workplace but also for those being trained in managing such projects. Indeed, the use of simulations has been quite common in the training of information systems professionals [62]. Through this technology, learners can see the impact of their decisions and explore possibilities before working with real-world projects where mistakes may have significant and far-reaching implications for any respective stakeholders. Ruiz et al. [51] reinforces this sentiment when they write, "simulation models offer, as a main advantage, the possibility of experimenting with different management decisions", continuing, "thus, it becomes possible to analyze the effect of those decisions in systems where the cost or risks of experimentation make it unfeasible" [51]. The implication here is that future projects will have access to better-trained project managers, which will optimally result in a higher percentage of successful projects.

*6.2. Limitations and Next Steps*

There are five limitations of this work that should be acknowledged and articulated as they also concurrently offer direction for next steps in the future development of this applied research. The first four are directly connected to the study's narrow scope. Our first limitation arises from our focus on simulating the waterfall model. While our solution offers a promising degree of adaptability, it might be beneficial for our community to examine how other models—such as parallel, iterative, and incremental models, among others—could be implemented as event-driven simulations in Python using SimPy. The next limitation relates to the exclusive use of Python to code our program. Future efforts could potentially explore the use of other languages, such as Java, via the Java Simulation Library (JSL), to extend the reach of our work. The third limitation lies in our adoption of an event-based method of simulation. Subsequent studies could explore alternative approaches, such as agent-based simulation. The fourth has to do with the technique that we employed to identify a resource level that would eliminate the time spent by phases in waiting for resources, a simple step algorithm. Exploring alternate optimization techniques, as well as other goals (e.g., eliminating resources or idle time), or examining them would also be of interest. Finally, our fifth limitation recognizes that any software is intrinsically bound by the human element. The current iteration of our program requires one to be comfortable working with code, as it lacks a graphical user interface to facilitate user interaction. Extending the software to include such a feature should expand the user base. Furthermore, an exploration into the usability of the output and how it can be further developed would certainly be of value.

Our Python simulation system opens several avenues for further study. While the simulation identifies bottlenecks and suggests optimal resource allocation in our hypothetical projects, studying the simulation's use in the real world will allow us to determine its efficiency in helping teams deliver software projects on time and within budget. A second area of further study involves integrating the simulation into software engineering courses to see how asking students to interact with and study the parameters and code of the simulation alters the ways in which they understand the software engineering process and the software development lifecycle.

**Computer Code and Software:** The simulation software can be downloaded from the following GitHub repository: https://github.com/adelphi-ed-tech/waterfall-sym.

**Author Contributions:** Conceptualization, A.S. and M.X.C.; methodology, A.S. and M.X.C.; software, A.S. and M.X.C.; writing—original draft preparation, A.S. and M.X.C.; writing—review and editing, A.S. and M.X.C. All authors have read and agreed to the published version of this manuscript.

**Funding:** This research was funded in part through the New York University School of Professional Studies Full-Time Faculty Professional Development Fund.

**Institutional Review Board Statement:** Not applicable.

**Informed Consent Statement:** Not applicable.

**Data Availability Statement:** Not applicable.

**Conflicts of Interest:** The authors declare no conflicts of interest.